\documentclass[prl,aps,twocolumn,preprintnumbers,showpacs,superscriptaddress]{revtex4}

\usepackage{amsfonts,amssymb,amsmath}            
\usepackage{graphics,graphicx,epsfig}            
\usepackage{amsthm}                              

\newcommand{\ie}{i.e.}
\newcommand{\eg}{e.g.}
\newcommand{\XX}{$XY$\ }  

\newcommand{\ket}[1]{\left | \, #1 \right\rangle}
\newcommand{\bra}[1]{\left \langle #1 \, \right |}

\newcommand{\half}{\mbox{$\textstyle \frac{1}{2}$}}
\newcommand{\h}[1]{\mathcal{H}_{#1} }

\newcommand{\be}{\begin{equation}}
\newcommand{\ee}{\end{equation}}
\newcommand{\bea}{\begin{eqnarray}}
\newcommand{\eea}{\end{eqnarray}}

\begin{document}


\title{Perfect state transfer in quantum spin networks}

\date[]{May 5, 2004}

\author{Matthias \surname{Christandl}}
\email[]{matthias.christandl@qubit.org}
\affiliation{Centre for Quantum
Computation,
             Centre for Mathematical Sciences,
             DAMTP,
             University of Cambridge,
             Wilberforce Road,
             Cambridge CB3 0WA, UK}

\author{Nilanjana \surname{Datta}}
\affiliation{Statistical Laboratory,
             Centre for Mathematical Science,
             University of Cambridge,
             Wilberforce Road,
             Cambridge CB3 0WB, UK }

\author{Artur \surname{Ekert}}
\affiliation{Centre for Quantum Computation,
             Centre for Mathematical Sciences,
             DAMTP,
             University of Cambridge,
             Wilberforce Road,
             Cambridge CB3 0WA, UK}
\affiliation{Department of Physics,
             National University of Singapore,
             Singapore 117\,542, Singapore}

\author{Andrew J. \surname{Landahl}}
\affiliation{Center for Bits and Atoms,
             Massachusetts Institute of Technology,
             Cambridge, MA 02139, USA}
\affiliation{HP Labs,
             Palo Alto, CA 94304-1126, USA}


\begin{abstract}

We propose a class of qubit networks that admit perfect transfer of any
quantum state in a fixed period of time. Unlike many other schemes for
quantum computation and communication, these networks do not require qubit
couplings to be switched on and off.  When restricted to $N$-qubit spin
networks of identical qubit couplings, we show that $2\log_3 N$ is the
maximal perfect communication distance for hypercube geometries. Moreover, if
one allows fixed but different couplings between the qubits then perfect
state transfer can be achieved over arbitrarily long distances in a linear
chain.

\end{abstract}



\maketitle 


The transfer of quantum states from one location (A) to another (B) is an
important feature in many quantum information processing systems. Depending
on technology at hand, this task can be accomplished in a number of ways.
Optical systems, typically employed in quantum communication and cryptography
applications, transfer states from $A$ to $B$ directly via photons. These
photons could contain an actual message or could be used to create
entanglement between $A$ and $B$ for future quantum teleportation between the
two sites \cite{Bennett:1992a}. Quantum computing applications with trapped
atoms use a variety of information carriers to transfer states from $A$ to
$B$, e.g. photons in cavity QED \cite{Guthorlein:2001} and phonons in ion
traps \cite{Leibfried:2003}. These photons and phonons may be viewed as
individual quantum carriers. However, many promising technologies for the
implementation of quantum information processing, such as optical
lattices~\cite{Mandel:2003}, and arrays of
quantum dots~\cite{dots} rely on collective phenomena to transfer quantum
states. In this case a ``quantum wire'', the most fundamental unit of any
quantum processing device, is made out of many interacting components. In the
sequel we focus on quantum channels of this type. Insight into the physics of
perfect quantum channels is of special significance for technologies that
route entanglement and quantum states on networks. These technologies range
from the very small, like the components of a quantum cellular automaton, to
the medium-sized, like the data bus of a quantum computer, to the truly
grand, like a quantum Internet spanning many quantum computers.

In this Letter, we address the problem of arranging $N$ interacting qubits in
a network which allows the perfect transfer of any quantum state over the
longest possible distance. The transfer is implemented by preparing the input
qubit $A$ in a prescribed quantum state and, some time later, by retrieving
the state from the output qubit $B$. The network is described by a graph $G$
in which the vertices $V(G)$ represent locations of the qubits and a set of
edges $E(G)$ specifies which pairs of qubits are coupled. The graph is
characterized by its adjacency matrix $A(G)$,
\be%
A_{ij}(G):=
\begin{cases}
1\mbox{ if } (i,j) \in E(G)\\
0\mbox{ otherwise.}
\end{cases}
\label{eq:adjacency}
\ee%
It has two special vertices, labelled as $A$ and $B$, which mark the input
and the output qubits respectively. We define the distance between $A$ and
$B$ to be the number of edges constituting the shortest path between them.
Although this distance is defined on a graph, it is directly related to the
physical separation between the input and output qubits, when a graph can be
embedded in physical space.

The most desirable graph, for our purposes, is a linear chain of $N$ qubits
with $A$ and $B$ at the two opposite ends of the chain. For fixed $N$ it
maximizes the distance between $A$ and $B$. If we can switch on and off
couplings between adjacent qubits then we can swap qubit states one-by-one
along the chain, all the way from the input to the output. Such a dynamical
control over the interactions between the qubits is still an experimental
challenge. Theoretically, this challenge has been met by considerable
progress in reducing the amount of control needed to accomplish quantum
computation tasks~\cite{Lloyd:1993a}. Moreover, it has been shown that this
can also be achieved without direct control over inter-qubit interactions, as
long as one has control over individual qubits~\cite{Lloyd:2001a}.
 Even if
just one qubit in the chain is controllable, then quantum communication can
be effected~\cite{Lloyd:2003a}.

Quantum communication over short distances through an unmodulated spin chain
has been studied in detail, and an expression for the fidelity of quantum
state has been obtained~\cite{Bose:2003}. In contrast to these works, we
focus on the situation in which the state transfer is \emph{perfect}, i.e.\
the fidelity is unity, and can be achieved over arbitrarily long distances.

We consider networks in which state transfer is achieved by time--evolution
under a suitable time--independent Hamiltonian, without any additional
external control. This mechanism avoids possible errors arising
from dynamical control of inter--qubit interactions.
However, note that we do not consider the effects of any other source of
errors in this paper.

Here, we show that a simple \XX coupling
\be%
H_G=\frac{1}{2} \sum_{(i,j) \in E(G)} \Bigl[ \sigma_i^x \sigma_j^x+\sigma_i^y
\sigma_j^y \Bigr], \label{eq:xxham}
\ee%
where $\sigma_i^x, \sigma_i^y$ and $\sigma_i^z $ are the Pauli matrices
acting on the $i\,$th qubit, allows perfect state transfer between antipodes
of a hypercube. Moreover, if we can engineer the strength of the couplings
between the qubits then perfect state transfer can be also performed between
the two ends of a linear chain. These tasks can be also accomplished with the
Heisenberg, or exchange, interaction by a suitable modulation of the network,
\eg, by placing it in a static but, in general, non-uniform external magnetic
field. We shall elaborate on this point later on.

Although our qubits represent generic two state systems, for convenience of
exposition we will also use the term spin as it provides a simple physical
picture of the network. The standard basis for an individual qubit is chosen
to be $\{\ket{0}\equiv\ket{\downarrow}, \,\ket{1}\equiv\ket{\uparrow}\}$ and
we shall assume that initially all spins point `down' along a prescribed
$z$-axis, \ie\ the network is in the state $\ket{\underline{0}}= \ket{0_A00
\cdots 00_B}$. This is an eigenstate of the Hamiltonian~(\ref{eq:xxham})
corresponding to zero energy.

The Hilbert space ${\mathcal{H}_{G}}$ associated with a network of $N$ qubits
is of dimension $2^N$.  However, the state transfer dynamics is completely
determined by the evolution in the $N$-dimensional subspace  ${\cal S}_G$
spanned by the basis vectors $\ket{n}$, $n=1, \ldots, N$, corresponding to
spin configurations in which all spins are `down' apart from just one spin at
the vertex $n$ which is `up'. Indeed, when we prepare the input qubit $A$ in
state $\alpha\ket{0}+\beta\ket{1}$, the state of the network becomes
\be%
\alpha\ket{0_A00 \cdots 00_B} +\beta\ket{1_A00 \cdots
00_B}=\alpha\ket{\underline{0}}+\beta\ket{1}. \label{eq:input}
\ee%
The coefficient $\alpha$ does not change in time, as $\ket{\underline{0}}$ is
the zero-energy eigenstate of $H_G$. The operator of the total $z$-component
of the spin,
\be%
\sigma_{tot}^z:= \sum_{i \in V(G)}\sigma_i^z,
\label{eq:spintot}%
\ee%
commutes with $H_G$, which leads to the conservation of the total
$z$-component of spin. This means that the state $\ket{1}\equiv\ket{1_A00
\cdots 00_B}$ must evolve into a superposition of states with exactly one
spin `up' and all other spins `down'. Thus the initial state of the network
evolves in time $t$ as
\be%
\alpha \ket{\underline{0}} +
\beta\ket{1}\mapsto\alpha\ket{\underline{0}}+\sum_{n=1}^N
\beta_n(t) \ket{n}.\label{eq:evolution}
\ee%
The dynamics are effectively confined to the subspace ${\cal S}_G$.  The
Hamiltonian $H_G$, when restricted to this subspace, is represented by an $N
\times N$ matrix that is \emph{identical} to the adjacency matrix $A(G)$,
Eq.~(\ref{eq:adjacency}), of the underlying graph $G$.  Due to this, one may
express the time evolution of the network in the ${\cal S}_G$ subspace as a
continuous-time quantum walk on $G$ (first considered by Farhi and Gutmann in
1998 \cite{Farhi:1998a}).

The question we are interested in is: When will the quantum walk propagate
from $A$ to $B$ with unit fidelity?  To answer this, if we identify qubit $A$
with vertex $1$ and qubit $B$ with vertex $N$, we need to compute the
probability amplitude that the network initially in state $\ket{1}$,
corresponding to $\ket{1_A00 \cdots 00_B}$, evolves after time $t$ to state
$\ket{N}$, corresponding to $\ket{0_A00 \cdots 01_B}$, \ie,
\be%
F(t)=\bra{N}e^{-i t H_G}\ket{1}.
\ee%
Perfect state transfer is obtained for times $t$ for which $|F(t)|=1$. Here
and henceforth we take $\hbar=1$.

Let us start with the \XX linear chain of qubits. In this case one can
compute $F(t)$ explicitly by diagonalizing the Hamiltonian or the
corresponding adjacency matrix. The eigenstates are given by
\bea
\tilde{\ket{k}} = \sqrt{\frac{2}{N+1}} \sum_{n=1}^N \sin\left(\frac{\pi k
n}{N+1}\right)\,\ket{n}
\label{eq:eigstates}%
\eea%
with corresponding eigenvalues $E_k  =  - 2\, \cos \frac{k \pi}{N+1}$ for all
$k=1, \ldots, N$. Thus
\be%
F(t)=\frac{2}{N+1}\sum_{k=1}^N \sin\Big(\frac{\pi k }{N+1}\Big) \sin
\Big(\frac{\pi k N}{N+1}\Big)e^{-i E_k t}.
\ee%

Perfect state transfer from one end of the chain to another is possible
\emph{only} for $N=2$ and $N=3$, with $F(t)=-i\sin(t)$ and
$F(t)=-\left[\sin\left(\frac{t}{\sqrt{2}}\right) \right]^2$ respectively. For
perfect state transfer in a chain, it is necessary that the ratios of the
differences of eigenvalues of the related adjacency matrix $A(G)$ are
rational numbers. The absence of perfect state transfer for $N \geq 4$ can be
proved by showing explicitly that the above condition is not satisfied.

A chain of two or three qubits can serve as basic building blocks for
networks that can perfectly transfer a quantum state over longer distances.
This can be achieved by building networks which are multiple Cartesian
products of either of the two simple chains.

In general the Cartesian product of two graphs $G:=\{V(G), E(G)\}$ and
$H:=\{V(H), E(H)\}$ is a graph $G \times H$ whose vertex set is $V(G) \times
V(H)$ and two of its vertices $(g,h)$ and $(g',h')$ are adjacent if and only
if one of the following hold: (i) $g=g'$ and $\{h,h'\} \in E(H)$; (ii) $h=h'$
and $\{g,g'\} \in E(G)$. If
$\tilde{\ket{k}}$ is an eigenvector of $A(G)$ corresponding to eigenvalue
$E_k$ and $\tilde{\ket{l}}$ is an eigenvector of $A(H)$ corresponding to
eigenvalue $E_l$ then $\tilde{\ket{k}}\otimes\tilde{\ket{l}}$ is an
eigenvector of $A(G\times H)$ corresponding to eigenvalue $E_k+E_l$. This is
because
\be%
A(G\times H) = A(G) \otimes \openone_{V(H)}  + \openone_{V(G)} \otimes A(H),
\label{eq:tp}
\ee%
where $\openone_{V(H)}$ is the $|V(H)| \times |V(H)|$ identity matrix (see
e.g. \cite{graphtheory}).

Now, consider a graph $G^d$ which is a $d$-fold Cartesian product of graph
$G$. The propagator between the two antipodal vertices in $G^d$, namely
$A=(1,\ldots,1)$ and $B=(N,\ldots,N)$, is simply
\be%
F_{G^d}(t)= \left[F_G (t)\right]^d.
\ee%
The $d$-fold Cartesian product of a one-link chain (two qubits) and a
two-link chain (three qubits) lead to one-link and two-link hypercubes with
$|F(t)|$ given, respectively, by
\be%
\left[\sin(t)\right]^d \quad\mbox{and}\quad
\left[\sin\left(\frac{t}{\sqrt{2}}\right) \right]^{2d}.
\ee%
Any quantum state can be perfectly transferred between the two antipodes of
the one-link and two-link hypercubes of any dimensions in constant time
$t=\pi/2$ and  $t=\pi/\sqrt{2}$ respectively~\footnote{Note that continuous
time quantum walk on the 1-link hypercube has recently been studied by
C.~Moore and A.~Russell in \emph{Randomization and Approximation Techniques:
6th International Workshop, (RANDOM 2002), Lecture Notes in Computer
Science}, (Springer-Verlag, Heidelberg), vol. 2483, p. 164.}.

Let us mention in passing that our discussion here is related to comparative
studies of continuous-time random walks on graphs. The mean hitting time
between vertices $A$ and $B$ is the time it takes the random walk on average
to reach $B$ starting at $A$. The classical mean hitting time between the
antipodes in a one-link and two-link $d$-dimensional hypercube is given, for
large $d$, by $2^d$ and $3^d$, respectively. One way to calculate this is to
reduce the continuous-time random walk on the $d$-dimensional hypercube $G^d$
to a continuous-time random walk, with potential drift, in \emph{one} and
\emph{two} dimensions, respectively, via the so-called lumping method (see
e.g.~\cite{Bovier:2003a}). In contrast, as we have shown, the corresponding
quantum hitting time is constant. This gives an exponential separation
between the classical and quantum mean hitting times between the antipodes of
one- and two-link hypercubes.

Thus we have shown that for a two-link hypercube of $N$ sites, the maximum
distance of perfect quantum communication is $2 \log_3 N$. It is an
interesting open problem to see if, given $N$ qubits, one can construct a
network with identical couplings in which any quantum state can be perfectly
transferred over a larger distance.

An improvement of the perfect quantum communication distance to $N$ is,
however, possible if one allows for different, but fixed, couplings between
qubits on a chain. In order to see how to do this, let us start with a
convenient relabelling of qubits. One may associate a fictitious
spin-$(N-1)/2$ particle with an $N$-qubit chain and relabel the basis vectors
as $\ket{m}$, where $m=-\half( N-1)+n-1$.

\begin{center}
\setlength{\unitlength}{0.85mm}
\begin{figure}
\begin{picture}(100,27)
\put(20,20){\line(1,0){75}}

\put(20,20){\circle*{2}}

\put(35,20){\circle*{2}}

\put(50,20){\circle*{2}}

\put(65,20){\circle*{2}}

\put(80,20){\circle*{2}}

\put(95,20){\circle*{2}}

 \put(0.5,22){\makebox(21,6){$J_n=$}}
 \put(17,22){\makebox(21,6){$\sqrt{1\cdot 5}$}}
 \put(32,22){\makebox(21,6){$\sqrt{2\cdot 4}$}}
 \put(47,22){\makebox(21,6){$\sqrt{3\cdot 3}$}}
 \put(62,22){\makebox(21,6){$\sqrt{4\cdot 2}$}}
 \put(77,22){\makebox(21,6){$\sqrt{5\cdot 1}$}}

 \put(7,12){\makebox(10,5){$n=$}}
 \put(15,12){\makebox(10,5){$1$}}
 \put(30,12){\makebox(10,5){$2$}}
 \put(45,12){\makebox(10,5){$3$}}
 \put(60,12){\makebox(10,5){$4$}}
 \put(75,12){\makebox(10,5){$5$}}
 \put(90,12){\makebox(10,5){$6$}}

 \put(6.5,6){\makebox(10,5){$m=$}}
 \put(13.5,6){\makebox(10,5){$-\frac{5}{2}$}}
 \put(28.5,6){\makebox(10,5){$-\frac{3}{2}$}}
 \put(43.5,6){\makebox(10,5){$-\frac{1}{2}$}}
 \put(60,6){\makebox(10,5){$\frac{1}{2}$}}
 \put(75,6){\makebox(10,5){$\frac{3}{2}$}}
 \put(90,6){\makebox(10,5){$\frac{5}{2}$}}
 \put(15,0){\makebox(10,5){$A$}}
 \put(90,0){\makebox(10,5){$B$}}
\end{picture}
\caption{Couplings $J_n$ that admit perfect state transfer from $A$ to $B$ in
a $6$-qubit chain. Eigenvalues $m$ of the equivalent spin-$\frac{5}{2}$
particle are also shown. }\label{fig:spinchain}
\end{figure}
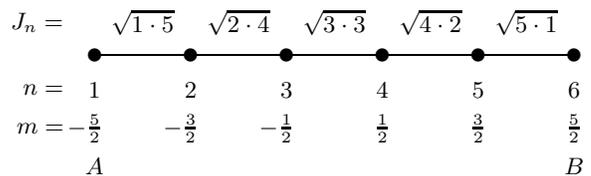
\end{center}
The input node $\ket{A}$ can be labelled both as $\ket{n=1}$ and
$\ket{m=-\half( N-1)}$, and the output node $\ket{B}$ both as $\ket{n=N}$ and
$\ket{m=+\half( N-1)}$.  An example for $N=6$ is depicted in
Fig.~\ref{fig:spinchain}. Now let the evolution of the chain be governed by a
modified version of~(\ref{eq:xxham}),
\be%
H_G=\sum_{(n,n+1) \in E(G)} \frac{J_n}{2}\Bigl[ \sigma_n^x
\sigma_{n+1}^x+\sigma_n^y \sigma_{n+1}^y \Bigr], \label{eq:hammod}
\ee%
which, when restricted to  the subspace ${\cal S}_G$, is of the form
\begin{equation}
\left(%
\begin{array}{ccccc}
  0      & J _1   & 0          & \cdots  & 0       \\
  J_1    & 0      & J_2        & \cdots  & 0       \\
  0      & J_2    & 0          & \cdots  & 0       \\
  \vdots & \vdots & \vdots    & \ddots  & J_{N-1} \\
  0      & 0      & 0            & J_{N-1} & 0       \\
\end{array}%
\right). \label{eq:hamhopp}
\end{equation}
The above matrix, Eq.~(\ref{eq:hamhopp}), is identical to the representation
of the Hamiltonian $H$ of a fictitious spin $S=\half (N-1)$ particle:
$H=\lambda S_x$, where $S_x$ is its angular momentum operator and $\lambda$
is some constant. In this case, the matrix elements $J_n$ are equal to
$\frac{\lambda}{2}\sqrt{n(N-n)}$. The evolution
\be%
U(t)=\exp \left(-i\lambda t\; S_x\right),
\ee%
of the network represents a rotation of this fictitious particle. The matrix
elements  $\bra{n'} U(t)\ket{n}$ of this rotation matrix are well-known and
in particular the probability amplitude for state transfer is
\be%
F(t)=\bra{N} U(t)\ket{1}=\left[- i \sin\left(\frac{\lambda
t}{2}\right)\right]^{N-1}. \label{eq:transferamplitude}
\ee%
Thus perfect transfer of a quantum state between the two antipodes $A$ and
$B$ is obtained in a constant time $t=\pi/\lambda$.

Each such engineered qubit chain can be viewed as a projection from a graph
having identical qubit couplings.  In fact, there is an entire family of such
graphs ${\cal G}$ that project to this chain.  Motivated by the `column
method' of \cite{Childs:2003a}, we define ${\cal G}$ as the set of graphs
whose vertices can be partitioned into $N$ columns $G_n$ of size
$|G_n|=\binom{N-1}{n-1}$ that satisfy the following two conditions for
$n=1,\ldots,N$: ($i$) each vertex in column $n$ is connected to $N-n$
vertices in column $n+1$, and ($ii$) each vertex in column $n+1$ is connected
to $n$ vertices in column $n$.  An important example of a graph in ${\cal G}$
is the one-link hypercube, where columns are defined as the set of vertices
reachable in $n$ links.  The evolution of a state at $A$ (the first column)
under $H_G$ (eq. (\ref{eq:xxham})) remains in the column space $\h{col}
\subseteq \h{G}$, spanned by
\be%
  \ket{\mbox{col } n}= \frac{1}{\sqrt{|G_n|}} \sum_{m=1}^{|G_n|}
  \ket{G_{n,m}}
\ee%
where $G_{n,m}$ labels the vertices in $G_n$. Hence, we restrict our
attention to $\h{col}$ in which the matrix elements of $H_G$ are given by
\be%
  J_n = \bra{\mbox{col } n} H_G \ket{\mbox{col } n+1} = \sqrt{n(N-n)},
\ee%
the same as in the engineered chain.

In our analysis we have focused on qubits coupled with the \XX interaction.
The choice of this interaction was dictated by its simple connection with the
adjacency matrix. We should add, however, that our considerations remain
valid if we choose the Heisenberg interaction and compensate for the diagonal
elements in the $S_G$ subspace. For example, the Heisenberg model with local
magnetic fields,
\be%
\half\sum_{j=1}^{N-1} J_j \;\vec{\sigma}_j\cdot \vec{\sigma}_{j+1}
+\sum_{j=1}^N B_j\sigma^{z}_j \label{eq:heisenberg2},
\ee%
where $\vec{\sigma_j}=(\sigma_j^x, \sigma_j^y, \sigma_j^x)$ and $B_n
=\frac{1}{2} (J_{n-1} + J_n ) - \frac{1}{2(N-2)} \sum_{k=1}^{N-1} J_k$, gives
exactly the same state transfer dynamics as the \XX model.

Our analysis is not restricted to pure states; the method presented here
works equally well for mixed states. It can also be used to transfer or to
distribute quantum entanglement.

In conclusion, in this Letter we have proven that perfect quantum state
transfer between antipodal points of one-link and two-link hypercubes is
possible and perfect quantum state transfer between antipodal points of
$N$-link hypercubes for $N\ge 3$ is impossible. The transfer time on these
hypercubes is independent of their dimension. In addition, we have shown that
a quantum state can be transferred perfectly over a chain of \emph{any}
length as long as one can pre-engineer inter-qubit interactions.  These
networks are especially appealing as they require no dynamical control,
unlike many other quantum communication proposals.

\begin{acknowledgments}
This work was supported by the CMI grant, A$^*$Star Grant No.\
012-104-0040 and the EU under project RESQ (IST-2001-37559). MC
was supported by a DAAD Doktorandenstipendium and the EPSRC. ND
would like to thank N.~Cooper and A.~Bovier for helpful
discussions. AL was supported by Hewlett-Packard and would like to
thank A.~Childs, S.~Gutmann, E.~Farhi, and J.~Goldstone for
collective helpful discussions.
\end{acknowledgments}


\end{document}